\documentclass[11pt]{article}
\pdfoutput=1  

\usepackage[mode=buildnew]{standalone}

\usepackage[T1]{fontenc}

\usepackage{microtype}

\usepackage{csquotes}

\newcommand{\emqu}[1]{\emph{\enquote{#1}}}
\usepackage{xspace}

\usepackage{enumitem}

\usepackage{amssymb}
\usepackage{amsmath}
\usepackage{complexity}
\usepackage{hyperref}
\usepackage[capitalize,noabbrev]{cleveref}
\let\Ptime\P
\newcommand{\pl}{\ensuremath{\scriptstyle\boxplus}\xspace}
\newcommand{\mi}{\ensuremath{\scriptstyle\boxminus}\xspace}

\usepackage{amthm}

\newcommand{\one}{\hbox{\texttt{1}}\xspace}
\newcommand{\zero}{\hbox{\texttt{0}}\xspace}
\newcommand{\cmos}{\textsc{cmos}\xspace}
\newcommand{\binput}{\textsc{input}\xspace}

\newcommand{\boutput}{\textsc{output}\xspace}

\newcommand{\bor}{\textsc{or}\xspace}

\newcommand{\band}{\textsc{and}\xspace}

\newcommand{\bnot}{\textsc{not}\xspace}

\newcommand{\nand}{\textsc{nand}\xspace}
\newcommand{\Nand}{\textsc{Nand}\xspace}
\newcommand{\xor}{\textsc{xor}\xspace}

\usepackage{bbold} 
\newcommand{\Nbb}{\mathbb{N}}
\newcommand{\Z}{\mathbb{Z}}

\usepackage{geometry}

\usepackage{subcaption}

\usepackage{tikz}
\usetikzlibrary{calc}

\usepackage{famacros}
\usepackage{fapics}

\newlength{\lenelresize}

\usepackage[draft]{fixme}
\fxusetheme{colorsig}
\FXRegisterAuthor{am}{anam}{AM}
\FXRegisterAuthor{tw}{atw}{TW}

\usepackage[
  backend=biber,bibencoding=utf8,sortcites=true,maxbibnames=99
]{biblatex}
\AtEveryBibitem{%
  \clearfield{doi}
  \clearfield{isbn}
  \clearfield{issn}
  \clearfield{note}
  \clearfield{url}
}

\title{Embedding arbitrary Boolean circuits into fungal automata}
\author{Augusto Modanese and Thomas Worsch \\
 Karlsruhe Institute of Technology (KIT), Karlsruhe, Germany}
\date{}

\begin{document}

\maketitle

\begin{abstract}
  Fungal automata are a variation of the two-dimensional sandpile automaton of
  \citeauthor{Bak_1987_SOC_prl} (\citefield{Bak_1987_SOC_prl}{journaltitle},
  \citeyear{Bak_1987_SOC_prl}).
  In each step toppling cells emit grains only to \emph{some} of their
  neighbors chosen according to a specific update sequence.
  We show how to embed any Boolean circuit into the initial configuration of a
  fungal automaton with update sequence $HV$.
  In particular we give a constructor that, given the description $B$ of a
  circuit, computes the states of all cells in the finite support of the
  embedding configuration in $O(\log |B|)$ space.
  As a consequence the prediction problem for fungal automata with
  update sequence $HV$ is $\Ptime$-complete.
  This solves an open problem of \citeauthor{Goles_CUF_2021_pla}
  (\citefield{Goles_CUF_2021_pla}{journaltitle}, \citeyear{Goles_CUF_2021_pla}).
\end{abstract}

\section{Introduction}

The two-dimensional sandpile automaton by
\citeauthor{Bak_1987_SOC_prl} \cite{Bak_1987_SOC_prl} has been
investigated from different points of view.
Because of the simple local rule, it is easily generalized to the
$d$-dimensional case for any integer $d\geq 1$.

Several \emph{prediction problems} for these cellular automata (CA) have been
considered in the literature.
Their difficulty varies with the dimensionality.
The recent survey by \citeauthor{Formenti_2020_HHPSLS_fi}
\cite{Formenti_2020_HHPSLS_fi} gives a good overview.
For one-dimensional sandpile CA the problems are known to be easy (see, e.g.,
\cite{Miltersen_2007_CCODS_tcs}).
For $d$-dimensional sandpile CA where $d\geq 3$, they are known to be
$\Ptime$-complete \cite{Moore_1999_CCS_jsp}.
In the two-dimensional case the situation is unclear; analogous
results are not known.

\emph{Fungal automata (FA)} as introduced by
\citeauthor{Goles_CUF_2021_pla} \cite{Goles_CUF_2021_pla} are a
variation of the two-dimensional sandpile automaton where a toppling
cell (i.e., a cell with state $\geq 4$) emits $2$ excess grains of sand either
to its two horizontal (\enquote{$H$}) or to its two vertical neighbors
(\enquote{$V$}).
These two modes of operation may alternate depending on an \emph{update
sequence} specifying in which steps grains are moved horizontally and in which
steps vertically.

The construction in \cite{Goles_CUF_2021_pla} shows that some natural
prediction problem is $\Ptime$-complete for two-dimensional fungal
automata with update sequence $H^4V^4$ (i.e., grains are first transferred
horizontally for $4$ steps and then vertically for $4$ steps, alternatingly).
The paper leaves open whether the same holds for shorter update sequences.
The shortest non-trivial sequence is $HV$ (and its complement $VH$); at the
same time this appears to be the most difficult to use.
By a reduction from the well-known \emph{circuit value problem} (CVP),
which is $\Ptime$-complete, we will show:

\begin{theorem}
  \label{main-thm}
  The following prediction problem is $\Ptime$-complete for FA with
  update sequence $HV$:
  
  \textbf{Given} as inputs initial states for a finite rectangle $R$
  of cells, a cell index $y$ (encoded in binary), and an upper bound
  $T$ (encoded in unary) on the number of steps of the FA,

  \textbf{decide} whether cell $x$ is in a state $\not=0$ or not at some time
  $t\le T$ when the FA is started with $R$ surounded by cells all in
  state $0$.
\end{theorem}
We assume readers are familiar with cellular automata (see \cref{sec_layer0} for
the definition).
We also assume knowledge of basic facts about Boolean circuits and complexity
theory, some of which we recall next.

\subsection{Boolean circuits and the CVP}
\label{subsec:circuits}

A Boolean circuit is a directed acyclic graph of \emph{gates}: \bnot gates
(with one input), \band and \bor gates with two inputs, $n\ge 1$ \binput
gates and one \boutput gate.
The output of a gate may be used by an arbitrary number of other gates.
Since a circuit is a dag and each gate obtains its inputs from gates in previous
layers, ultimately the output of each gate can be computed from a subset of the
input gates in a straightforward way.

It is straightforward to realize \bnot, \band, and \bor gates in terms of \nand
gates with two inputs (with an only constant overhead in the number of gates).
To simplify the construction later on, we assume that circuits consist
exclusively of \nand gates.

Each gate of a circuit is described by a $4$-tuple $(g,t,g_1,g_2)$
where $g$ is the number of the gate, $t$ describes the type of the
gate, and $g_1$ and $g_2$ are the numbers of the gates (called sources
of $g$) which produce the inputs for gate $g$; all numbers are
represented in binary.
If gate $g$ has only one input, then $g_2=g_1$ by convention.
Without loss of generality the \binput gates have numbers $1$ to $n$
and since their predecessors $g_1$ and $g_2$ will never be used,
assume they are set to $0$.
All other gates have subsequent numbers starting at $n+1$ such
that the inputs for gate $g$ are coming from gates with strictly
smaller numbers.
Following \textcite{Ruzzo_1981_UCC_ar} the description $B$ of a
complete circuit is the concatenation of the descriptions of all of
its gates, sorted by increasing gate numbers.

Problem instances of the \emph{circuit value problem} (CVP) consist of the
description $B$ of a Boolean circuit $C$ with $n$ inputs and a list
$x$ of $n$ input bits.
The task is to decide whether $C(x) = 1$ holds or not.
It is well known that the CVP is $\Ptime$-complete.

\subsection{Challenges}

A standard strategy for showing $\Ptime$-completeness of a problem $\Pi$ in some
computational model $\mathcal{M}$ (and also the one employed by
\citeauthor{Goles_CUF_2021_pla} in \cite{Goles_CUF_2021_pla}) is by a reduction
from the CVP to $\Pi$, which entails describing how to \enquote{embed} circuits
in $\mathcal{M}$.

In our setting of fungal automata with update sequence $HV$, while realizing
wires and signals as in \cite{Goles_CUF_2021_pla} is possible, there is no
obvious implementation for negation nor for a reliable wire crossing.
Hence, it seems one can only directly construct circuits that are \emph{both}
planar and monotone.
Although it is known that the CVP is $\Ptime$-complete for \emph{either} planar
or monotone circuits \cite{Goldschlager_1977_MPCVP_sigactn}, it is unlikely that
one can achieve the same under both constraints.
This is because the CVP for circuits that are both monotone and planar lies in
$\NC^2$ (and is thus certainly not $\Ptime$-complete unless $\Ptime \subseteq
\NC^2$) \cite{Dymond_HCPC_1980_focs}.

We are able to overcome this barrier by exploiting features that are present in
fungal automata but not in general circuits: \emph{time} and \emph{space}.
Namely, we deliberately \emph{retard} signals in the circuits we implement by
extending the length of the wires that carry them.
We show how this allows us to realize a primitive form of transistor.
From this, in turn, we are able to construct a \nand gate, thus allowing both
wire crossings and negations to be implemented.

Our construction is not subject to the limitations that apply to the
two-dimensional case that were previously shown by
\citeauthor{Gajardo_2006_CITS_tcs} in \cite{Gajardo_2006_CITS_tcs} since the FA
starting configuration is not a fixed point.
The resulting construction is also significantly more complex than that of
\cite{Goles_CUF_2021_pla}.

\subsection{Overview of the construction}

%
In the rest of the paper we describe how to embed any Boolean circuit
with description $B$ and an assignment of values to the inputs into a
configuration $c$ of a fungal automaton in such a way that the
following holds:
\begin{itemize}
\item \enquote{Running} the FA for a sufficient number of steps
  results in the \enquote{evaluation} of all simulated gates.
  In particular, after reaching a stable configuration, a specific
  cell of the FA is in state $1$ or $0$ if and only if the output of
  the circuit is $1$ or $0$, respectively.
\item The initial configuration $F$ of the FA is simple in the sense that, given
  the description of a circuit and an input to it, we can produce its embedding
  $F$ using $O(\log n + \log |B|)$ space.
  Thus we have a log-space reduction from the CVP to the prediction
  problem for FA.
\end{itemize}
The construction consists of several layers:
\begin{description}
\item[Layer $0$:] The underlying model of fungal automata.
\item[Layer $1$:] As a first abstraction we subdivide the space into
  \emqu{blocks} of $2\times2$ cells and always think of update \emqu{cycles}
  consisting of $4$ steps of the CA, using the update sequence $(HV)^2$.
\item[Layer $2$:] On top of that we will implement \emqu{polarized
    circuits} processing \emqu{polarized signals} that run along \emqu{wires}.
\item[Layer $3$:] Polarized circuitry is then used to implement
  \emqu{Boolean circuits with delay}: \emqu{bits} are processed by
  \emqu{gates} connected by \emqu{cables}.%
  \footnote{Here we slightly deviate from the standard terminology of Boolean
  circuits and reserve the term \emqu{wire} for the more primitive wires defined
  in layer 2.}
\item[Layer $4$:] Finally a given Boolean circuit (without delay) can be
embedded in a fungal automaton (as a circuit with delay) in a systematic fashion
that needs only logarithmic space to construct.
\end{description}
The rest of this paper has a simple organization:
Each layer $i$ will be described separately in section $i+2$.

\section{Layer 0: The Fungal Automaton}
\label{sec_layer0}

Let $\Nbb_+$ denote the set of positive integers and $\Z$ that of all integers.
For $d \in \Nbb_+$, a \emqu{$d$-dimensional CA} is a tuple $(S,N,\delta)$ where:
\begin{itemize}[nosep]
  \item $S$ is a finite set of states
  \item $N$ is a finite subset of $\Z^d$, called the \emqu{neighborhood}
  \item $\delta\colon S^N \to S$ is the \emqu{local transition function}
\end{itemize}
In the context of CA, the elements of $\Z^d$ are referred to as \emph{cells}.
The function $\delta$ induces a \emqu{global transition function} $\Delta\colon
S^{\Z^d} \to S^{\Z^d}$ by applying $\delta$ to each cell simultaneously.
In the following, we will be interested in the case $d = 2$ and the so-called
\emph{von Neumann neighborhood} $N = \{ (a,b) \in \Z^2 \mid |a|+|b| \le 1 \}$
of radius $1$.

Except for the updating of cells the fungal automaton is just a two-dimensional
CA with the von Neumann neighborhood of radius $1$ and $S=\{0,1,\dots,7\}$ as
the set of states.%
\footnote{We use states as in \cite{Bak_1987_SOC_prl}; however, the states $6$
and $7$ never occur in our construction.}
A \emqu{configuration} is thus a mapping $c: \Z^2 \to S$.

Depending on the their states cells will be depicted as follows in
diagrams:

\begin{tabular}[t]{l@{\qquad}l@{\qquad}l}
  -- state $0$ as \textcell{-}
  & -- state $1$ as \textcell{.}
  & -- state $i \in S \setminus \{0,1\}$ as \textcell{i}
\end{tabular}

We will use colored background for cells in states $2$, $3$, and $4$ since their
presence determines the behavior of the polarized circuit.
The state $1$ is only a \enquote{side effect} of an empty cell receiving a grain
of sand from some neighbors; hence it is represented as a dot.
Cells which are not included in a figure are always assumed to be in state $0$.

For a logical predicate $P$ denote by $[P]$ the value $1$ if $P$ is
true and the value $0$ if $P$ is false.
For $i\in \Z^2$ denote by $h(i)$ the two horizontal neighbors of cell
$i$ and by $v(i)$ its two vertical neighbors.
Cells are updated according to $2$ functions $H$ and $V$ mapping from
$S^{\Z^2}$ to $S^{\Z^2}$ where for each $i\in \Z^2$ the following holds:
\begin{align*}
  H(c)(i) &= c(i) - 2\cdot [c(i)\geq 4] + \sum_{j\in h(i)} [c(j)\geq 4]; \\
  V(c)(i) &= c(i) - 2\cdot [c(i)\geq 4] + \sum_{j\in v(i)} [c(j)\geq 4].
\end{align*}
The updates are similar to the sandpile model by
\citeauthor{Bak_1987_SOC_prl} \cite{Bak_1987_SOC_prl}, but toppling
cells only emit grains of sand either to their horizontal or their
vertical neighbors.
Therefore whenever a cell is non-zero, it stays non-zero forever.

\begin{figure}
  \centering
  \begin{tikzpicture}[fa]
  \begin{fasimple}[shift={(0,0)}]
    \farow{-------}
    \farow{-DCCCCC}
    \farow{-BCCCCC}
    \farow{-------}
  \end{fasimple}
  \drawH{R2C7.south east}

  \begin{fasimple}[shift={(8.5,0)}]
    \farow{-------}
    \farow{PBDCCCC}
    \farow{-BCCCCC}
    \farow{-------}
  \end{fasimple}
  \drawV{R2C7.south east}

  \begin{fasimple}[shift={(17,0)}]
    \farow{--P----}
    \farow{PBBCCCC}
    \farow{-BDCCCC}
    \farow{-------}
  \end{fasimple}
  \drawH{R2C7.south east}
  
  \begin{scope}[shift={(0,5)}]
    
    \begin{fasimple}[shift={(0,0)}]
      \farow{--P----}
      \farow{PBBCCCC}
      \farow{-CBDCCC}
      \farow{-------}
    \end{fasimple}
    \drawV{R2C7.south east}

    \begin{fasimple}[shift={(8.5,0)}]
      \farow{--P----}
      \farow{PBBDCCC}
      \farow{-CBBCCC}
      \farow{---P---}
    \end{fasimple}
    \drawH{R2C7.south east}

    \begin{fasimple}[shift={(17,0)}]
      \farow{--P----}
      \farow{PBCBDCC}
      \farow{-CBBCCC}
      \farow{---P---}
    \end{fasimple}
  \end{scope}



\end{tikzpicture}
  \caption{Five transitions according to $HVHVH$}
  \label{fig:fa-wire-cycle}
\end{figure}
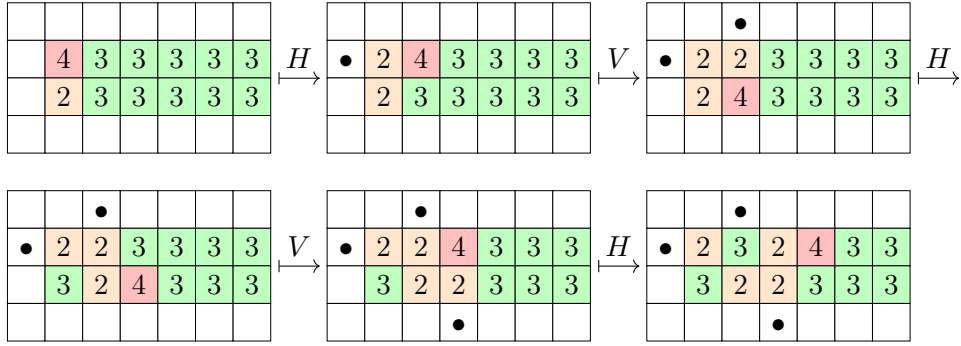

The composition of these functions applying first $H$ and then $V$ is
denoted $HV$.
For the transitions of a fungal automaton with update sequence $HV$
these functions are applied alternatingly, resulting in a computation
$c$, $H(c)$, $V(H(c))$, $H(V(H(c)))$, $V(H(V(H(c))))$, and so on. In
examples we will often skip three intermediate configurations and only
show $c$, $HVHV(c)$, etc.
\cref{fig:fa-wire-cycle} shows a simple first example.

\section{Layer 1: Coarse Graining Space and Time}
\label{sec_layer1}

\begin{figure}
  \centering
  \begin{tikzpicture}[fa]
    \begin{farows}[shift={(0,0)}]
      \farow{DCCCCC}
      \farow{BCCCCC}
    \end{farows}
    \drawHVHV{R1C6.south east}

    \begin{farows}[shift={(7.5,0)}]
      \farow{BBDCCC}
      \farow{CBBCCC}
    \end{farows}
    \drawHVHV{R1C6.south east}

    \begin{farows}[shift={(15,0)}]
      \farow{BCBBDC}
      \farow{CBCBBC}
    \end{farows}
\end{tikzpicture}
  \caption{compact representation of two cycles}
  \label{fig:fa-wire-cycle-compact}
\end{figure}
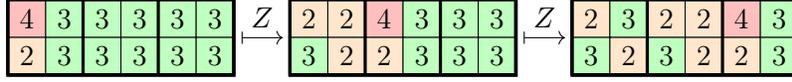

As a first abstraction from now on one should always think of the
space as subdivided into \emqu{blocks} of $2\times2$ cells.
Furthermore we will look at update \emqu{cycles} consisting of $4$
steps of the CA, thus using the update sequence $HVHV$ which we will
abbreviate to $Z$.
As an example \Cref{fig:fa-wire-cycle-compact} shows the same cycle as
\cref{fig:fa-wire-cycle} and the following cycle in a compact way.
Block boundaries are indicated by thicker lines.

Cells outside the depicted area of a figure are assumed to be
\textcell{0} initially and they will never become critical and topple
during the shown computation.

\section{Layer 2: Polarized Components}
\label{sec_layer2}

We turn to the second lowest level of abstraction.
Here we work with two types of signals, which we refer to as \emph{positive}
(denoted \pl) and \emph{negative} (denoted \mi).
Both types will have several representations as a block in the FA.
\begin{itemize}
\item All representations of a \pl signal have in common that the
  \emph{\textbf{upper} left corner} of the block is a \textcell{4} and
  the other cells are \textcell{2} or \textcell{3}.
\item All representations of a \mi signal have in common that the
  \emph{\textbf{lower} left corner} of the block is a \textcell{4} and
  the other cells are \textcell{2} or \textcell{3}.
\end{itemize}
Not all representations will be appropriate in all situations as will
be discussed in the next subsection.

The rules of fungal automata allow us to perform a few basic operations on
these \emph{polarized} signals (e.g., duplicating, merging, or crossing them
under certain assumptions).
The highlight here is that we can implement a (delay-sensitive) form of
transistor that works with polarized signals, which we refer to as a
\emph{switch}.

As a convention, in the figures in this section, we write $x$ and $y$ for the
inputs of a component and $z$, $z_1$, and $z_2$ for the outputs.

\subsection{Polarized Signals and Wires}

\begin{figure}
  \centering
  \hspace*{\fill}
  \begin{subfigure}[t]{0.4\linewidth}
    \centering
      \begin{tikzpicture}[fa]
    \farowsetup
    \farow{DC}
    \farow{CC}
    \draw[fa/grid] (R1C1.north west) rectangle (R2C2.south east);
  \end{tikzpicture}
  \hspace{2mm}
  \begin{tikzpicture}[fa]
    \farowsetup
    \farow{DC}
    \farow{BC}
    \draw[fa/grid] (R1C1.north west) rectangle (R2C2.south east);
  \end{tikzpicture}
  \hspace{2mm}
  \begin{tikzpicture}[fa]
    \farowsetup
    \farow{DC}
    \farow{BB}
    \draw[fa/grid] (R1C1.north west) rectangle (R2C2.south east);
  \end{tikzpicture}
  \hspace{2mm}
  \begin{tikzpicture}[fa]
    \farowsetup
    \farow{DB}
    \farow{BB}
    \draw[fa/grid] (R1C1.north west) rectangle (R2C2.south east);
  \end{tikzpicture}
    \caption{\pl signals}
    \label{fig_pl_signals}
  \end{subfigure}
  \hspace*{\fill}
  \begin{subfigure}[t]{0.4\linewidth}
    \centering
      \begin{tikzpicture}[fa]
    \farowsetup
    \farow{CC}
    \farow{DC}
    \draw[fa/grid] (R1C1.north west) rectangle (R2C2.south east);
  \end{tikzpicture}
  \hspace{2mm}
  \begin{tikzpicture}[fa]
    \farowsetup
    \farow{BC}
    \farow{DC}
    \draw[fa/grid] (R1C1.north west) rectangle (R2C2.south east);
  \end{tikzpicture}
  \hspace{2mm}
  \begin{tikzpicture}[fa]
    \farowsetup
    \farow{BB}
    \farow{DC}
    \draw[fa/grid] (R1C1.north west) rectangle (R2C2.south east);
  \end{tikzpicture}
  \hspace{2mm}
  \begin{tikzpicture}[fa]
    \farowsetup
    \farow{BB}
    \farow{DB}
    \draw[fa/grid] (R1C1.north west) rectangle (R2C2.south east);
  \end{tikzpicture}
    \caption{\mi signals}
    \label{fig_mi_signals}
  \end{subfigure}
  \hspace*{\fill}
  \caption{Representations of \pl and \mi signals}
  \label{fig_pl_mi_signals}
\end{figure}
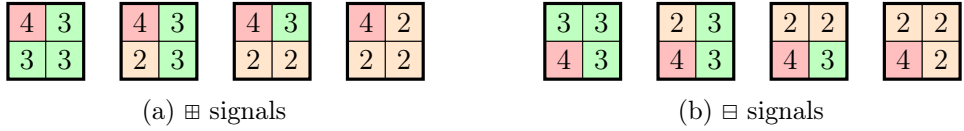

Representations of \pl and \mi signals are shown in \cref{fig_pl_mi_signals}.
We will refer to a block initially containing a \pl or \mi signal as a \pl or
\mi \emph{source}, respectively.
(This will be used, for instance, to set the inputs to the embedded CVP
instance.)

A comparison of \cref{fig:fa-wire-cycle-compact} and \cref{fig_pl_signals} shows
that in the former a \pl signal is \enquote{moving from left to right}.
In general we will use \emph{wires} to propagate signals. 
Wires extending horizontally or vertically can be constructed by
juxtaposing \emph{wire blocks} consisting of $2\times2$ blocks of cells in state
\textcell{3}.

While one can use the same wire blocks for both types of signals, each block is
destroyed upon use and thus can only be used once.
In particular, this means a wire will either be used by a \pl or a \mi signal.
We refer to the respective wires as \emph{\pl} and \emph{\mi wires},
accordingly.

\begin{figure}
  \centering
  \resizebox{\textwidth}{!}{%
      \begin{tikzpicture}[fa]
    \begin{scope}[shift={(0,0)}]
      \farowsetup
      \farow{--CCCCCC}
      \farow{--CCCCCC}
      \farow{--CC--CC}
      \farow{--CC--CC}
      \farow{DCCC--CC}
      \farow{BCCC--CC}
      \foreach \r in {1, 3, 5} {\draw[fa/grid] (R\r C1.north west) -- (R\r C8.north east);}
      \draw[fa/grid] (R6C1.south west) -- (R6C8.south east);
      \foreach \c in {1, 3, 5, 7} {\draw[fa/grid] (R1C\c.north west) -- (R6C\c.south west);}
      \draw[fa/grid] (R1C8.north east) -- (R6C8.south east);
      \draw[fa/grid] (R1C1.north west) rectangle (R6C8.south east);
    \end{scope}
    \drawHVHV{R3C8.south east}
    
    \begin{scope}[shift={(9.5,0)}]
      \farowsetup
      \farow{--CCCCCC}
      \farow{--CCCCCC}
      \farow{--CC--CC}
      \farow{--CC--CC}
      \farow{BBDC--CC}
      \farow{CBBC--CC}
      \foreach \r in {1, 3, 5} {\draw[fa/grid] (R\r C1.north west) -- (R\r C8.north east);}
      \draw[fa/grid] (R6C1.south west) -- (R6C8.south east);
      \foreach \c in {1, 3, 5, 7} {\draw[fa/grid] (R1C\c.north west) -- (R6C\c.south west);}
      \draw[fa/grid] (R1C8.north east) -- (R6C8.south east);
      \draw[fa/grid] (R1C1.north west) rectangle (R6C8.south east);
    \end{scope}
    \drawHVHV{R3C8.south east}
    
    \begin{scope}[shift={(19,0)}]
      \farowsetup
      \farow{--CCCCCC}
      \farow{--CCCCCC}
      \farow{--DC--CC}
      \farow{--BB--CC}
      \farow{BCCB--CC}
      \farow{CBCB--CC}
      \foreach \r in {1, 3, 5} {\draw[fa/grid] (R\r C1.north west) -- (R\r C8.north east);}
      \draw[fa/grid] (R6C1.south west) -- (R6C8.south east);
      \foreach \c in {1, 3, 5, 7} {\draw[fa/grid] (R1C\c.north west) -- (R6C\c.south west);}
      \draw[fa/grid] (R1C8.north east) -- (R6C8.south east);
      \draw[fa/grid] (R1C1.north west) rectangle (R6C8.south east);
    \end{scope}    
    \drawHVHV{R3C8.south east}

    \begin{scope}[shift={(0,7.5)}]
      \farowsetup
      \farow{--DCDCCC}
      \farow{--BBBCCC}
      \farow{--CB--CC}
      \farow{--BC--CC}
      \farow{BCCB--CC}
      \farow{CBCB--CC}
      \foreach \r in {1, 3, 5} {\draw[fa/grid] (R\r C1.north west) -- (R\r C8.north east);}
      \draw[fa/grid] (R6C1.south west) -- (R6C8.south east);
      \foreach \c in {1, 3, 5, 7} {\draw[fa/grid] (R1C\c.north west) -- (R6C\c.south west);}
      \draw[fa/grid] (R1C8.north east) -- (R6C8.south east);
      \draw[fa/grid] (R1C1.north west) rectangle (R6C8.south east);
    \end{scope}
    \drawHVHV{R3C8.south east}

    \begin{scope}[shift={(9.5,7.5)}]
      \farowsetup
      \farow{--BCBBDC}
      \farow{--BCCBBC}
      \farow{--CB--DC}
      \farow{--BC--CC}
      \farow{BCCB--CC}
      \farow{CBCB--CC}
      \foreach \r in {1, 3, 5} {\draw[fa/grid] (R\r C1.north west) -- (R\r C8.north east);}
      \draw[fa/grid] (R6C1.south west) -- (R6C8.south east);
      \foreach \c in {1, 3, 5, 7} {\draw[fa/grid] (R1C\c.north west) -- (R6C\c.south west);}
      \draw[fa/grid] (R1C8.north east) -- (R6C8.south east);
      \draw[fa/grid] (R1C1.north west) rectangle (R6C8.south east);
    \end{scope}
    \drawHVHV{R3C8.south east}

    \begin{scope}[shift={(19,7.5)}]
      \farowsetup
      \farow{--BCBBBB}
      \farow{--BCCBCC}
      \farow{--CB--CB}
      \farow{--BC--BB}
      \farow{BCCB--DC}
      \farow{CBCB--CC}
      \foreach \r in {1, 3, 5} {\draw[fa/grid] (R\r C1.north west) -- (R\r C8.north east);}
      \draw[fa/grid] (R6C1.south west) -- (R6C8.south east);
      \foreach \c in {1, 3, 5, 7} {\draw[fa/grid] (R1C\c.north west) -- (R6C\c.south west);}
      \draw[fa/grid] (R1C8.north east) -- (R6C8.south east);
      \draw[fa/grid] (R1C1.north west) rectangle (R6C8.south east);
    \end{scope}
  \end{tikzpicture}
  }
  \caption{A \pl signal moving along a wire with two right turns.}
  \label{fig:meander-upper-half}
\end{figure}
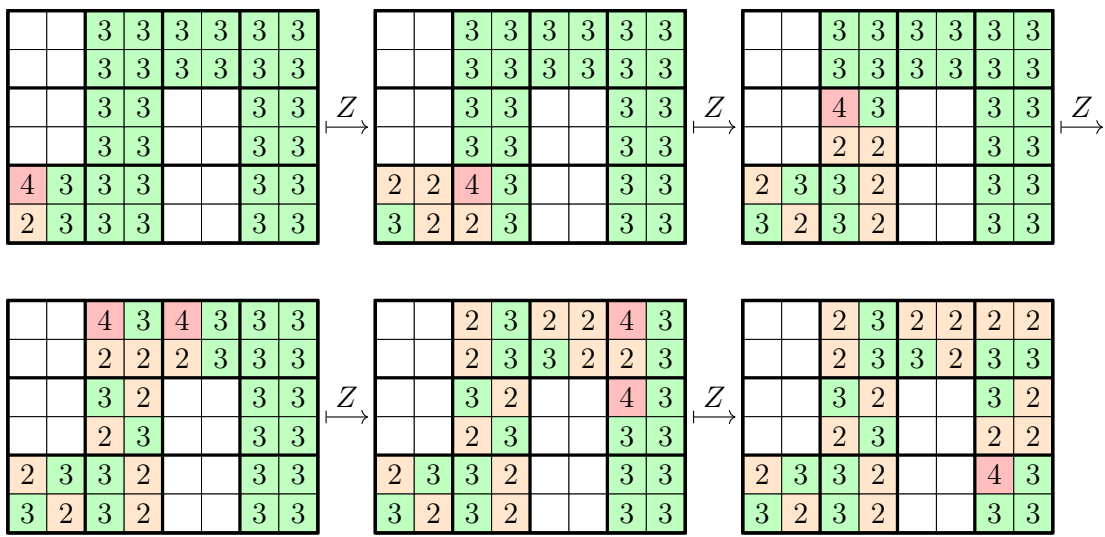

Every representation of a signal is restricted with respect to the possible
directions it can move to along a wire.
In our construction each signal will start at the left end of a
horizontal wire.
\cref{fig:meander-upper-half} shows how a \pl signal first \enquote{turns left}
once and then moves along a wire that \enquote{turns right} two times, changing
its representation while meandering around.
(The case of a \mi signal is similar and is not shown.)

\cref{fig:meander-lower-half} can be seen as the continuation of
\cref{fig:meander-upper-half}. The \pl signal moves further down, 
\enquote{turns left} twice, and then reaches the end of the wire.
The composition of both parts can be seen in \cref{fig_retarder} and
will be used as the basic building unit for \enquote{retarders}.


\begin{figure}
  \centering
  \resizebox{\textwidth}{!}{%
      \begin{tikzpicture}[fa]
    \begin{scope}[shift={(0,0)}]
      \farowsetup
      \farow{DC--CCCC}
      \farow{CC--CCCC}
      \farow{CC--CC--}
      \farow{CC--CC--}
      \farow{CCCCCC--}
      \farow{CCCCCC--}
      \foreach \r in {1, 3, 5} {\draw[fa/grid] (R\r C1.north west) -- (R\r C8.north east);}
      \draw[fa/grid] (R6C1.south west) -- (R6C8.south east);
      \foreach \c in {1, 3, 5, 7} {\draw[fa/grid] (R1C\c.north west) -- (R6C\c.south west);}
      \draw[fa/grid] (R1C8.north east) -- (R6C8.south east);
      \draw[fa/grid] (R1C1.north west) rectangle (R6C8.south east);
    \end{scope}
    \drawHVHV{R3C8.south east}

    \begin{scope}[shift={(9.5,0)}]
      \farowsetup
      \farow{CB--CCCC}
      \farow{BB--CCCC}
      \farow{DC--CC--}
      \farow{CC--CC--}
      \farow{CCCCCC--}
      \farow{CCCCCC--}
      \foreach \r in {1, 3, 5} {\draw[fa/grid] (R\r C1.north west) -- (R\r C8.north east);}
      \draw[fa/grid] (R6C1.south west) -- (R6C8.south east);
      \foreach \c in {1, 3, 5, 7} {\draw[fa/grid] (R1C\c.north west) -- (R6C\c.south west);}
      \draw[fa/grid] (R1C8.north east) -- (R6C8.south east);
      \draw[fa/grid] (R1C1.north west) rectangle (R6C8.south east);
    \end{scope}
    \drawHVHV{R3C8.south east}

    \begin{scope}[shift={(19,0)}]
      \farowsetup
      \farow{CB--CCCC}
      \farow{BC--CCCC}
      \farow{CB--CC--}
      \farow{BB--CC--}
      \farow{DCCCCC--}
      \farow{CCCCCC--}
      \foreach \r in {1, 3, 5} {\draw[fa/grid] (R\r C1.north west) -- (R\r C8.north east);}
      \draw[fa/grid] (R6C1.south west) -- (R6C8.south east);
      \foreach \c in {1, 3, 5, 7} {\draw[fa/grid] (R1C\c.north west) -- (R6C\c.south west);}
      \draw[fa/grid] (R1C8.north east) -- (R6C8.south east);
      \draw[fa/grid] (R1C1.north west) rectangle (R6C8.south east);
    \end{scope}
    \drawHVHV{R3C8.south east}

    \begin{scope}[shift={(0,7.5)}]
      \farowsetup
      \farow{CB--CCCC}
      \farow{BC--CCCC}
      \farow{CB--CC--}
      \farow{BC--CC--}
      \farow{CBDCCC--}
      \farow{BBBCCC--}
      \foreach \r in {1, 3, 5} {\draw[fa/grid] (R\r C1.north west) -- (R\r C8.north east);}
      \draw[fa/grid] (R6C1.south west) -- (R6C8.south east);
      \foreach \c in {1, 3, 5, 7} {\draw[fa/grid] (R1C\c.north west) -- (R6C\c.south west);}
      \draw[fa/grid] (R1C8.north east) -- (R6C8.south east);
      \draw[fa/grid] (R1C1.north west) rectangle (R6C8.south east);
    \end{scope}
    \drawHVHV{R3C8.south east}

    \begin{scope}[shift={(9.5,7.5)}]
      \farowsetup
      \farow{CB--CCCC}
      \farow{BC--CCCC}
      \farow{CB--CC--}
      \farow{BC--CC--}
      \farow{CCBBDC--}
      \farow{BBCBBC--}
      \foreach \r in {1, 3, 5} {\draw[fa/grid] (R\r C1.north west) -- (R\r C8.north east);}
      \draw[fa/grid] (R6C1.south west) -- (R6C8.south east);
      \foreach \c in {1, 3, 5, 7} {\draw[fa/grid] (R1C\c.north west) -- (R6C\c.south west);}
      \draw[fa/grid] (R1C8.north east) -- (R6C8.south east);
      \draw[fa/grid] (R1C1.north west) rectangle (R6C8.south east);
    \end{scope}
    \drawHVHV{R3C8.south east}

    \begin{scope}[shift={(19,7.5)}]
      \farowsetup
      \farow{CB--CCCC}
      \farow{BC--CCCC}
      \farow{CB--DC--}
      \farow{BC--BB--}
      \farow{CCBBCB--}
      \farow{BBCBCB--}
      \foreach \r in {1, 3, 5} {\draw[fa/grid] (R\r C1.north west) -- (R\r C8.north east);}
      \draw[fa/grid] (R6C1.south west) -- (R6C8.south east);
      \foreach \c in {1, 3, 5, 7} {\draw[fa/grid] (R1C\c.north west) -- (R6C\c.south west);}
      \draw[fa/grid] (R1C8.north east) -- (R6C8.south east);
      \draw[fa/grid] (R1C1.north west) rectangle (R6C8.south east);
    \end{scope}
    \drawHVHV{R3C8.south east}

    \begin{scope}[shift={(0,15)}]
      \farowsetup
      \farow{CB--DCDC}
      \farow{BC--BBBC}
      \farow{CB--CB--}
      \farow{BC--BC--}
      \farow{CCBBCB--}
      \farow{BBCBCB--}
      \foreach \r in {1, 3, 5} {\draw[fa/grid] (R\r C1.north west) -- (R\r C8.north east);}
      \draw[fa/grid] (R6C1.south west) -- (R6C8.south east);
      \foreach \c in {1, 3, 5, 7} {\draw[fa/grid] (R1C\c.north west) -- (R6C\c.south west);}
      \draw[fa/grid] (R1C8.north east) -- (R6C8.south east);
      \draw[fa/grid] (R1C1.north west) rectangle (R6C8.south east);
    \end{scope}
  \end{tikzpicture}
  }
  \caption{A \pl signal moving along with two left turns (continuation
    of \cref{fig:meander-upper-half})}
  \label{fig:meander-lower-half}
\end{figure}




\subsection{Diodes}

Note that \pl and \mi signals do not encode any form of direction in them
(regarding their propagation along a wire).
In fact, a signal propagates in any direction a wire is placed in.
In order for our components to operate correctly, it will be necessary to ensure
a signal is propagated in a single direction.
To realize this, we use \emph{diodes}.

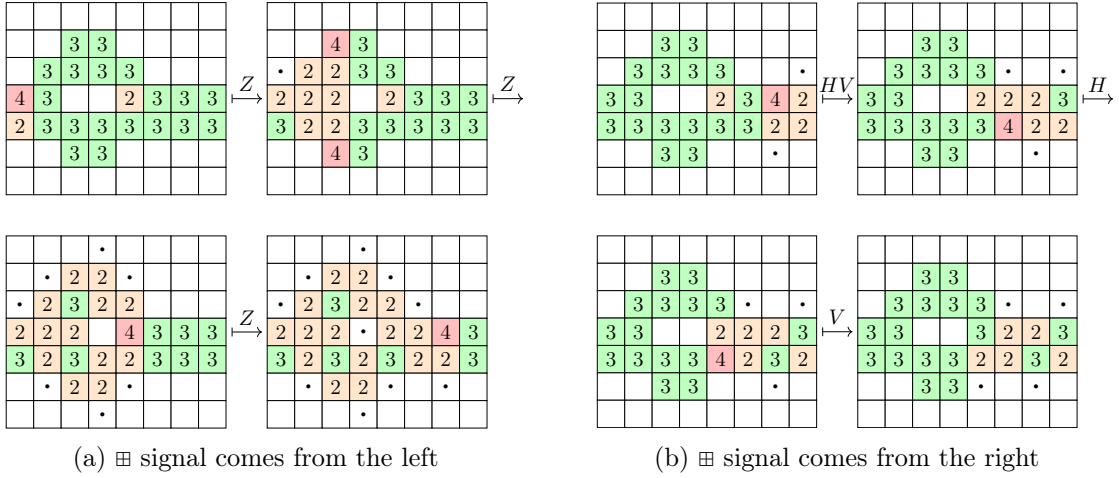
\begin{figure}
  \centering
  \begin{subfigure}{0.48\textwidth}
    \centering
    \resizebox{\textwidth}{!}{%
      \begin{tikzpicture}[fa]
  \begin{fasimple}[shift={(0,0)}]
    \farow{--------}
    \farow{--CC----}
    \farow{-CCCC---}
    \farow{DC--BCCC}
    \farow{BCCCCCCC}
    \farow{--CC----}
    \farow{--------}
  \end{fasimple}
  \drawHVHV{R4C8.east}

  \begin{fasimple}[shift={(9.5,0)}]
    \farow{--------}
    \farow{--DC----}
    \farow{.BBCC---}
    \farow{BBB-BCCC}
    \farow{CBBCCCCC}
    \farow{--DC----}
    \farow{--------}
  \end{fasimple}
  \drawHVHV{R4C8.east}

  \begin{fasimple}[shift={(0,8.5)}]
    \farow{---.----}
    \farow{-.BB.---}
    \farow{.BCBB---}
    \farow{BBB-DCCC}
    \farow{CBCBBCCC}
    \farow{-.BB.---}
    \farow{---.----}
  \end{fasimple}
  \drawHVHV{R4C8.east}

  \begin{fasimple}[shift={(9.5,8.5)}]
    \farow{---.----}
    \farow{-.BB.---}
    \farow{.BCBB.--}
    \farow{BBB.BBDC}
    \farow{CBCBCBBC}
    \farow{-.BB.-.-}
    \farow{---.----}
  \end{fasimple}
\end{tikzpicture}
    }
    \caption{\pl signal comes from the left}
  \end{subfigure}
  \hfill
  \begin{subfigure}{0.48\textwidth}
    \centering
    \resizebox{\textwidth}{!}{%
      \begin{tikzpicture}[fa]
  \begin{fasimple}[shift={(0,0)}]
    \farow{--------}
    \farow{--CC----}
    \farow{-CCCC--.}
    \farow{CC--BCDB}
    \farow{CCCCCCBB}
    \farow{--CC--.-}
    \farow{--------}
  \end{fasimple}
  \drawHV{R4C8.east}


  \begin{fasimple}[shift={(9.5,0)}]
    \farow{--------}
    \farow{--CC----}
    \farow{-CCCC.-.}
    \farow{CC--BBBC}
    \farow{CCCCCDBB}
    \farow{--CC--.-}
    \farow{--------}
  \end{fasimple}
  \drawH{R4C8.east}

  \begin{fasimple}[shift={(0,8.5)}]
    \farow{--------}
    \farow{--CC----}
    \farow{-CCCC.-.}
    \farow{CC--BBBC}
    \farow{CCCCDBCB}
    \farow{--CC--.-}
    \farow{--------}
  \end{fasimple}
  \drawV{R4C8.east}

  \begin{fasimple}[shift={(9.5,8.5)}]
    \farow{--------}
    \farow{--CC----}
    \farow{-CCCC.-.}
    \farow{CC--CBBC}
    \farow{CCCCBBCB}
    \farow{--CC.-.-}
    \farow{--------}
  \end{fasimple}
\end{tikzpicture}
    }
    \caption{\pl signal comes from the right}
  \end{subfigure}
  \caption{Diode operation on \pl wires}
  \label{fig_diode_sim}
\end{figure}

A diode is an element on a horizontal wire that only allows a signal to flow
from left to right.
A signal coming from right to left is not allowed through.
As the other components, the diode is intended to be used only once.
For the implementation, refer to \cref{fig_diode_impl}.
(Recall that $x$ denotes the component's input and $z$ its output.)
\begin{figure}
  \centering
  \hspace*{\fill}
  \begin{subfigure}[t]{0.3\textwidth}
    \centering
    \begin{tikzpicture}[fa]
  \begin{farows}
    \farow{--CC----}
    \farow{-CCCC---}
    \farow{XE--BCZE}
    \farow{EECCCCEE}
    \farow{--CC----}
    \farow{--------}
  \end{farows}
\end{tikzpicture}
    \caption{\pl wires}
  \end{subfigure}
  \hspace*{\fill}
  \begin{subfigure}[t]{0.3\textwidth}
    \centering
    \begin{tikzpicture}[fa]
    \begin{farows}
      \farow{--------}
      \farow{--CC----}
      \farow{XECCCCZE}
      \farow{EE--BCEE}
      \farow{-CCCC---}
      \farow{--CC----}
    \end{farows}
\end{tikzpicture}
    \caption{\mi wires}
  \end{subfigure}
  \hspace*{\fill}
  \caption{Diode implementations}
  \label{fig_diode_impl}
\end{figure}
\cref{fig_diode_sim} illustrates the operation of a diode for \pl signals.
(The case of \mi signals is similar.)

For all the remaining elements described in this section, we implicitly add
diodes to their inputs and outputs.
This ensures that the signals can only flow from left to right (as intended).
This is probably not necessary for all elements, but doing so makes the
construction simpler while the overhead is only a constant factor blowup in the
size of the elements.

\subsection{Duplicating, Merging, and Crossing Wires}

\begin{figure}
  \centering
  \hspace*{\fill}
  \begin{subfigure}[t]{0.3\textwidth}
    \centering
    \begin{tikzpicture}[fa]
  \begin{farows}
    \farow{----CCUE}
    \farow{---CCCEE}
    \farow{XECC----}
    \farow{EECCC---}
    \farow{----CCVE}
    \farow{-----CEE}
  \end{farows}
\end{tikzpicture}
    \caption{Duplicating \pl wires}
  \end{subfigure}
  \hspace*{\fill}
  \begin{subfigure}[t]{0.3\textwidth}
    \centering
    \begin{tikzpicture}[fa]
    \begin{farows}
      \farow{-----CUE}
      \farow{----CCEE}
      \farow{XECCC---}
      \farow{EECC----}
      \farow{---CCCVE}
      \farow{----CCEE}
    \end{farows}
\end{tikzpicture}
    \caption{Duplicating \mi wires}
  \end{subfigure}
  \hspace*{\fill}
  \\[3mm]
  \hspace*{\fill}
  \begin{subfigure}[t]{0.3\textwidth}
    \centering
    \begin{tikzpicture}[fa]
  \begin{farows}
    \farow{XECC----}
    \farow{EECCC---}
    \farow{----CCZE}
    \farow{---CCCEE}
    \farow{YECC----}
    \farow{EEC-----}
  \end{farows}
\end{tikzpicture}
    \caption{Merging \pl wires}
  \end{subfigure}
  \hspace*{\fill}
  \begin{subfigure}[t]{0.3\textwidth}
    \centering
    \begin{tikzpicture}[fa]
    \begin{farows}
      \farow{XEC-----}
      \farow{EECC----}
      \farow{---CCCZE}
      \farow{----CCEE}
      \farow{YECCC---}
      \farow{EECC----}
    \end{farows}
\end{tikzpicture}
    \caption{Merging \mi wires}
  \end{subfigure}
  \hspace*{\fill}
  \caption{Duplicating and merging wires}
  \label{fig_dupl_merge_impl}
\end{figure}
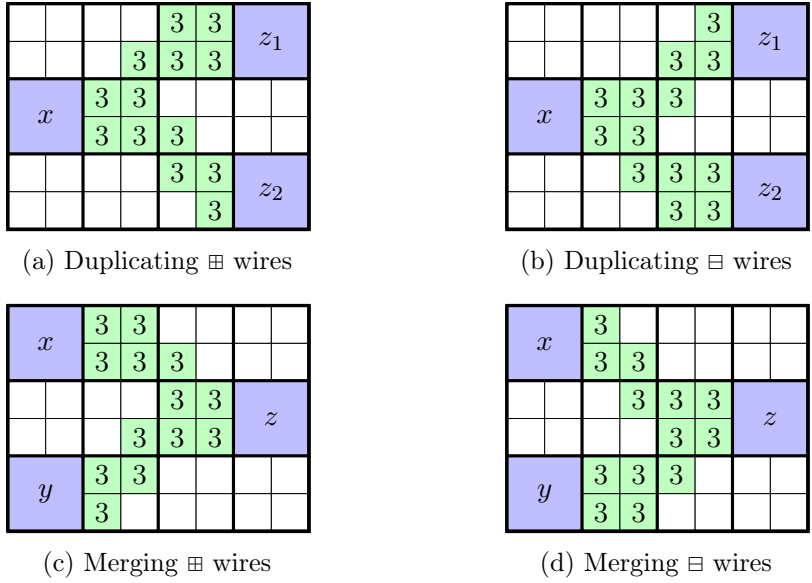

Wires of the same polarity can be \emph{duplicated} or \emph{merged}.
By duplicating a wire we mean we create two wires $z_1$ and $z_2$ from a single
wire $x$ in such a way that, if any signal arrives from $x$, then this signal is
duplicated and propagated on both $z_1$ and $z_2$.
(Equivalently, one might imagine that $x = z_1$ and $z_2$ is a wire copy of
$x$.)
In turn, a wire merge realizes in some sense the reverse operation:
We have two wires $x$ and $y$ \emph{of the same polarity} and create a wire $z$
such that, if a signal arrives from $x$ or $y$ (or both), then a signal of the
same polarity will emerge at $z$.
(Hence one could say the wire merge realizes a polarized \bor gate.)
See \cref{fig_dupl_merge_impl} for the implementations.

As discussed in the introduction, there is no straightforward realization of
a wire crossing in fungal automata in the traditional sense.
Nevertheless, it turns out we \emph{can} cross wires under the following
constraints:
\begin{enumerate}
  \item The two wires being crossed are a \pl and a \mi wire.
  \item The crossing is used only once and by a single input wire; that is, once
  a signal from either wire passes through the crossing, it is destroyed.
  (If two signals arrive from both wires at the same time, then the crossing is
  destroyed without allowing any signal to pass through.)
\end{enumerate}
To elicit these limitations, we refer to such crossings as \emph{semicrossings}.

\begin{figure}
  \centering
  \hspace*{\fill}
  \begin{subfigure}[t]{0.3\textwidth}
    \centering
    \begin{tikzpicture}[fa]
  \begin{farows}
    \farow{XE---CUE}
    \farow{EEC-CCEE}
    \farow{--CCC---}
    \farow{--CCC---}
    \farow{YEC-CCVE}
    \farow{EE---CEE}
  \end{farows}
\end{tikzpicture}
    \caption{\pl semicrossing}
  \end{subfigure}
  \hspace*{\fill}
  \begin{subfigure}[t]{0.3\textwidth}
    \centering
    \begin{tikzpicture}[fa]
    \begin{farows}
      \farow{XEC---UE}
      \farow{EECC-CEE}
      \farow{---CCC--}
      \farow{---CCC--}
      \farow{YECC-CVE}
      \farow{EEC---EE}
    \end{farows}
\end{tikzpicture}
    \caption{\mi semicrossing}
  \end{subfigure}
  \hspace*{\fill}
  \caption{Semicrossing implementations}
  \label{fig_semix_impl}
\end{figure}
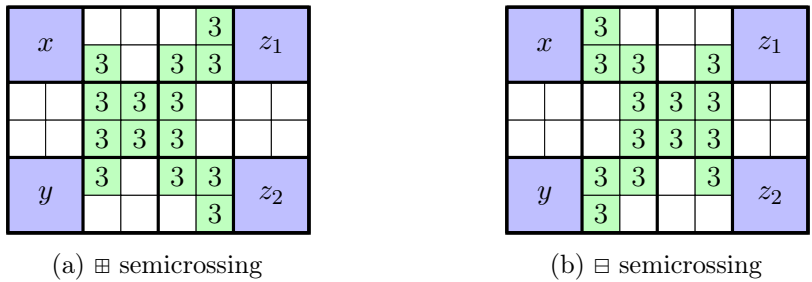

We actually need two types of semicrossings, one for each choice of polarities
for the two input wires.
The semicrossings are named according to the polarity of the top input wire:
A \pl semicrossing has a \pl wire as its top input (and a \mi wire as its
bottom one) whereas a \mi semicrossing has a \mi wire at the top (and a \pl wire
at the bottom).
For the implementations, see \cref{fig_semix_impl}.

\subsection{Switches}

A \emph{switch} is a rudimentary form of transistor.
It has two inputs and one output.
Adopting the terminology of field-effect transistors (FETs), we will refer to
the two inputs as the \emph{source} and \emph{gate} and the output as the
\emph{drain}.
In its initial state, the switch is \emph{open} and does not allow source signals
to pass through.
If a signal arrives from the gate, then it turns the switch \emph{closed}.
A subsequent signal arriving from the source will then be propagated on to the
drain.
This means that switches are \emph{delay-sensitive}:
A signal arriving at the source only continues on to the drain if the gate
signal has arrived beforehand (or simultaneously to the source).

\begin{figure}
  \centering
  \hspace*{\fill}
  \begin{subfigure}[t]{0.3\textwidth}
    \centering
    \begin{tikzpicture}[fa]
  \begin{farows}
    \farow{XE------}
    \farow{EECCC---}
    \farow{--CBCCZE}
    \farow{--CC-CEE}
    \farow{YEC-----}
    \farow{EE------}
  \end{farows}
\end{tikzpicture}
    \caption{\pl switch}
  \end{subfigure}
  \hspace*{\fill}
  \begin{subfigure}[t]{0.3\textwidth}
    \centering
    \begin{tikzpicture}[fa]
    \begin{farows}
      \farow{XE------}
      \farow{EEC-----}
      \farow{--CC-CZE}
      \farow{--CBCCEE}
      \farow{YECCC---}
      \farow{EE------}
    \end{farows}
\end{tikzpicture}
    \caption{\mi switch}
  \end{subfigure}
  \hspace*{\fill}
  \caption{Switch implementations}
  \label{fig_switch_impl}
\end{figure}
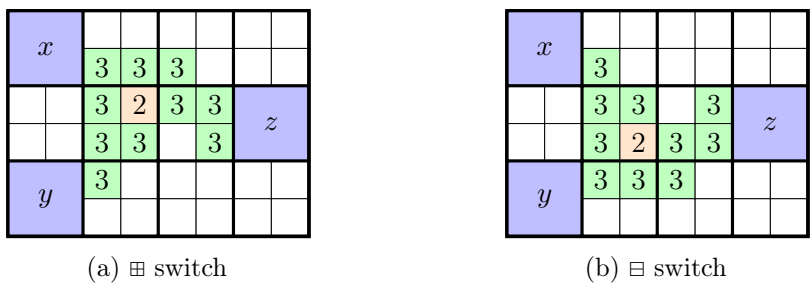

Similar to semicrossings, our switches come in two flavors.
In both cases the top input is a \pl wire and the bottom one a \mi.
The difference is that, in a \pl switch, the source (and thus also the drain) is
the \pl input and the gate is the \mi input.
Conversely, in a \mi switch the source and drain are \mi wires and the gate is a
\pl wire.
Refer to \cref{fig_switch_impl} for the implementation of the two types of
switches.

\subsection{Delays and Retarders}
\label{sec_delays}

As mentioned in the introduction, the circuits we construct are sensitive to the
time it takes for a signal to flow from one point to the other.
To render this notion precise, we define for every component a \emph{delay}
which results from the time taken for a signal to pass through the component.
This is defined as follows:
\begin{itemize}
  \item The delay of a source is zero.
  \item The delay of a wire (including bends) at some block $B$ is the delay of
  the wire's source $S$ plus the length (in blocks) of the shortest contiguous
  path along the wire that leads from $S$ to $B$ according to the von Neumann
  neighborhood.
  We will refer to this length as the \emph{wire distance} between $S$ and $B$.
  For example, the wire distance between the inputs and outputs in all of
  \cref{fig_diode_impl,fig_dupl_merge_impl,fig_switch_impl,fig_semix_impl} is
  $4$; similarly, the distance between $x$ and $z$ in \cref{fig_retarder} (see
  below) is $15$.
  \item The delay of a gate (i.e., a diode, wire duplication, wire merge, or
  semicrossing) is the maximum over the delays of its inputs plus the gate width
  (in blocks).
\end{itemize}
Notice our definition of wire distance may grossly estimate the actual number of
steps a signal requires to propagate from $S$ to $B$.
This is fine for our purposes since we only need to reason about upper bounds
later in \cref{sec_setting_delays}.

Finally we will also need a \emph{retarder} element, which is responsible for
adding a variable amount of delay to a wire.
Refer to \cref{fig_retarder} for their realization.
\begin{figure}
  \centering
    \begin{tikzpicture}[fa]
  \begin{farows}
    \farow{--CCCCCC--------}
    \farow{--CCCCCC--------}
    \farow{--CC--CC--------}
    \farow{--CC--CC--------}
    \farow{XECC--CC--CCCCZE}
    \farow{EECC--CC--CCCCEE}
    \farow{------CC--CC----}
    \farow{------CC--CC----}
    \farow{------CCCCCC----}
    \farow{------CCCCCC----}
    \farow{----------------}
  \end{farows}
\end{tikzpicture}
  \caption{Implementation of a basic retarder (for both \pl and \mi signals)
  that ensures a delay of $\ge 12$ at $z$ (relative to $x$).
  Retarders for greater delays can be realized by increasing (i) the height of
  the meanders, (ii) the number of up-down meanders, and (iii) the positions of
  the input and output.}
  \label{fig_retarder}
\end{figure}
Retarders can have different dimensions.
Evidently, one can ensure a delay of $t$ with a retarder that is $O(\sqrt{t})
\times O(\sqrt{t})$ large.
We are going to use retarders of delay at most $D$, where $D$ depends on the
CVP instance and is set later in \cref{sec_setting_delays}.
Hence, it is safe to assume all retarders in the same configuration are of the
same size horizontally and vertically, but realize different delays.
This allows one to use retarders of a \emph{single} size for any fixed circuit,
which simplifies the layout significantly (see also
\cref{sec_setting_delays,subsec:constructor}).

\section{Layer 3: Working With Bits}
\label{sec_layer3}

\begin{figure}
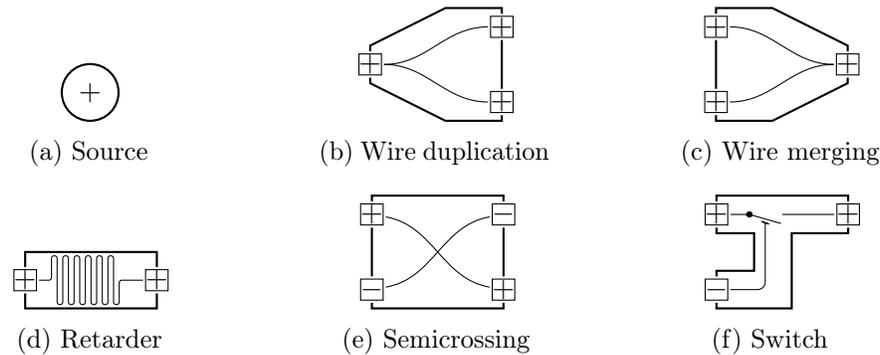

  \centering
  \hspace*{\fill}
  \begin{subfigure}[t]{0.2\textwidth}
    \centering
    \includestandalone{fig-el-layer3/source}
    \caption{Source}
  \end{subfigure}
  \hspace*{\fill}
  \begin{subfigure}[t]{0.2\textwidth}
    \centering
    \includestandalone{fig-el-layer3/dupl}
    \caption{Wire duplication}
  \end{subfigure}
  \hspace*{\fill}
  \begin{subfigure}[t]{0.2\textwidth}
    \centering
    \includestandalone{fig-el-layer3/merge}
    \caption{Wire merging}
  \end{subfigure}
  \hspace*{\fill}
  \\[3mm]
  \hspace*{\fill}
  \begin{subfigure}[t]{0.2\textwidth}
    \centering
    \includestandalone{fig-el-layer3/retarder}
    \caption{Retarder}
  \end{subfigure}
  \hspace*{\fill}
  \begin{subfigure}[t]{0.2\textwidth}
    \centering
    \includestandalone{fig-el-layer3/semix}
    \caption{Semicrossing}
  \end{subfigure}
  \hspace*{\fill}
  \hspace*{-3mm}  
  \begin{subfigure}[t]{0.2\textwidth}
    \centering
    \includestandalone{fig-el-layer3/switch}
    \caption{Switch}
  \end{subfigure}
  \hspace*{0mm}
  \hspace*{\fill}
  \caption{Representations of the elements from abstraction layer 2 as used in
  layer 3.
  The polarities indicate whether the \pl or \mi version of the component is used.}
  \label{fig_el_layer3}
\end{figure}

We will now use the elements from \cref{sec_layer2} (represented as in
\cref{fig_el_layer3}) to construct planar delay-sensitive Boolean circuits.
Our circuits will use \nand gates as their basis.
We discuss how to overcome the planarity restriction in \cref{sec_cable_x}.

\subsection{Representation of Bits}

For the representation of a bit, we use a pair consisting of a polarized \pl
wire and a polarized \mi wire.
Such a pair of polarized wires is called a \emph{cable}.
As mentioned earlier, most of the time signals will travel from left
to right.
It is straightforward to generalize the notion of wire distance (see
\cref{sec_delays}) to cables simply by setting it to the maximum of the
respective wire distances.

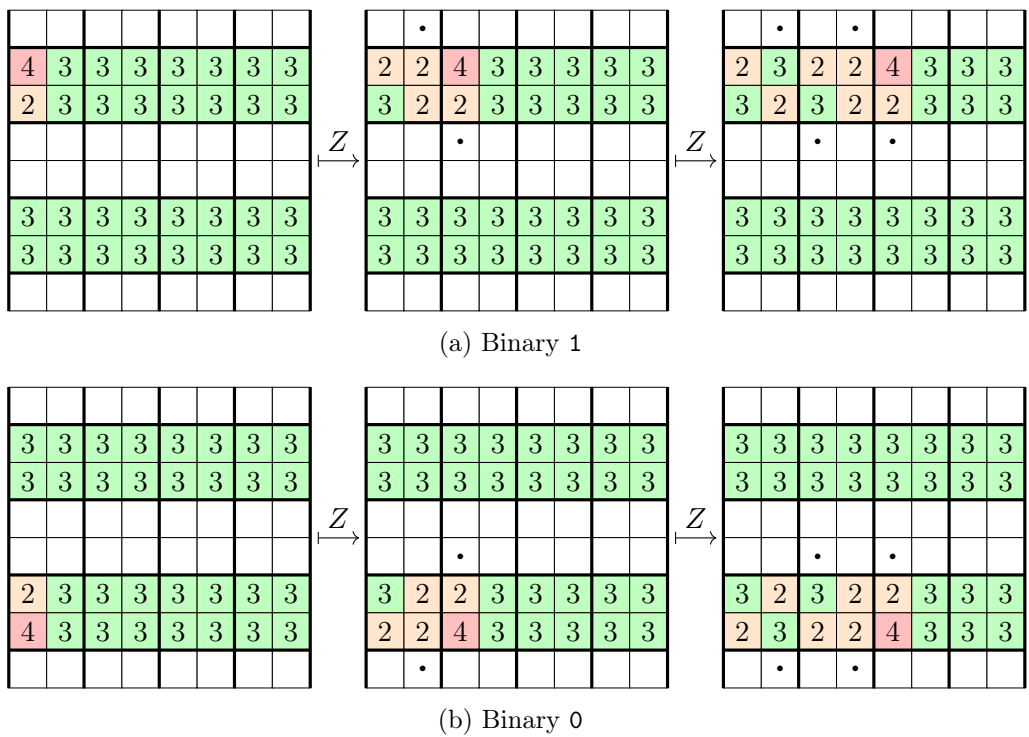
\begin{figure}
  \centering
  \begin{subfigure}{\textwidth}
    \centering
    \def\bitreprgrid{%
  \foreach \c in {1,3,5,7} {
    \draw[fa/grid] (R1C\c.north west) -- (R8C\c.south west);
  }
  \draw[fa/grid] (R1C8.north east) -- (R8C8.south east);
  \foreach \r in {2,4,6,8} {
    \draw[fa/grid] (R\r C1.north west) -- (R\r C8.north east);
  }
}
\begin{tikzpicture}[fa]
  \begin{scope}[shift={(0,0)}]
    \farowsetup
    \farow{--------}
    \farow{DCCCCCCC}
    \farow{BCCCCCCC}
    \farow{--------}
    \farow{--------}
    \farow{CCCCCCCC}
    \farow{CCCCCCCC}
    \farow{--------}
    \bitreprgrid
  \end{scope}
  \drawHVHV{R4C8.south east}

  \begin{scope}[shift={(9.5,0)}]
    \farowsetup
    \farow{-.------}
    \farow{BBDCCCCC}
    \farow{CBBCCCCC}
    \farow{--.-----}
    \farow{--------}
    \farow{CCCCCCCC}
    \farow{CCCCCCCC}
    \farow{--------}
    \bitreprgrid
  \end{scope}
  \drawHVHV{R4C8.south east}

  \begin{scope}[shift={(19,0)}]
    \farowsetup
    \farow{-.-.----}
    \farow{BCBBDCCC}
    \farow{CBCBBCCC}
    \farow{--.-.---}
    \farow{--------}
    \farow{CCCCCCCC}
    \farow{CCCCCCCC}
    \farow{--------}
    \bitreprgrid
  \end{scope}
\end{tikzpicture}%
    \caption{Binary \one}
  \end{subfigure}
  \\[3mm]
  \begin{subfigure}{\textwidth}
    \centering
    \def\bitreprgrid{%
  \foreach \c in {1,3,5,7} {
    \draw[fa/grid] (R1C\c.north west) -- (R8C\c.south west);
  }
  \draw[fa/grid] (R1C8.north east) -- (R8C8.south east);
  \foreach \r in {2,4,6,8} {
    \draw[fa/grid] (R\r C1.north west) -- (R\r C8.north east);
  }
}
\begin{tikzpicture}[fa]
  \begin{scope}[shift={(0,9.5)}]
    \begin{scope}[shift={(0,0)}]
      \farowsetup
      \farow{--------}
      \farow{CCCCCCCC}
      \farow{CCCCCCCC}
      \farow{--------}
      \farow{--------}
      \farow{BCCCCCCC}
      \farow{DCCCCCCC}
      \farow{--------}
      \bitreprgrid
    \end{scope}
    \drawHVHV{R4C8.south east}

    \begin{scope}[shift={(9.5,0)}]
      \farowsetup
      \farow{--------}
      \farow{CCCCCCCC}
      \farow{CCCCCCCC}
      \farow{--------}
      \farow{--.-----}
      \farow{CBBCCCCC}
      \farow{BBDCCCCC}
      \farow{-.------}
      \bitreprgrid
    \end{scope}
    \drawHVHV{R4C8.south east}

    \begin{scope}[shift={(19,0)}]
      \farowsetup
      \farow{--------}
      \farow{CCCCCCCC}
      \farow{CCCCCCCC}
      \farow{--------}
      \farow{--.-.---}
      \farow{CBCBBCCC}
      \farow{BCBBDCCC}
      \farow{-.-.----}
      \bitreprgrid
    \end{scope}
  \end{scope}
\end{tikzpicture}%
    \caption{Binary \zero}
  \end{subfigure}
  \caption{Binary representations travelling from left to right along a cable}
  \label{fig:bit-repr}
\end{figure}

A signal on a cable's \pl wire represents a binary \one, and a signal on the \mi
wire represents a binary \zero.
By convention we will always draw the \pl wire \enquote{above} the \mi
wire of the same cable.
See \cref{fig:bit-repr} for an example.

When referring to a gate's inputs and outputs, we indicate the \pl and \mi
components of a cable with subscripts.
For instance, for an input cable $x$, we write $x_+$ for its \pl and $x_-$ for
its \mi component.

\subsection{Bit Duplication}
\label{sec_bit_dupl}

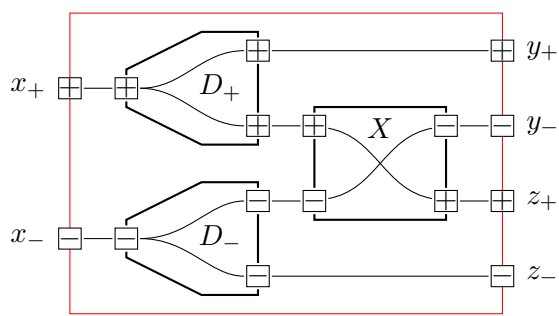
\begin{figure}
  \centering
    \begin{tikzpicture}[x={(5mm,0mm)}, y={(0mm,-5mm)}]
  \draw[red] (3.5,0) rectangle (15,8);

  \path (3.5,2) pic["$x_+$"] (xp) {I+};
  \path (3.5,6) pic["$x_-$"] (xm) {I-};



  \path (5,2) pic["$D_+$"] (B1p) {B+};
  \draw (xp/C) -- (B1p/Ip/C);

  \path (5,6) pic["$D_-$"] (B2m) {B-};
  \draw (xm/C) -- (B2m/Im/C);

  \path (10,3) pic["$X$"] (X) {X+};
  \draw (B1p/O2p/C) -- (X/Ip/C);
  \draw (B2m/O1m/C) -- (X/Im/C);

  \path (15,1) pic["$y_+$"] (yp) {O+};
  \path (15,3) pic["$y_-$"] (ym) {O-};
  \path (15,5) pic["$z_+$"] (zp) {O+};
  \path (15,7) pic["$z_-$"] (zm) {O-};
  \draw (B1p/O1p/C) -- (yp/C);
  \draw (X/Om/C) -- (ym/C);
  \draw (X/Op/C) -- (zp/C);
  \draw (B2m/O2m/C) -- (zm/C);
\end{tikzpicture}
  \caption{Boolean branch}
  \label{fig:bool-branch}
\end{figure}

To duplicate a cable, we use the \emph{Boolean branch} depicted in
\cref{fig:bool-branch}.
The circuit consists of two wire duplications (one of each polarity) and a
crossing.

\subsection{\Nand Gates}

As a matter of fact the \nand gate is inspired by the implementation
of such a gate in \cmos
technology\footnote{e.\,g.~\url{https://en.wikipedia.org/wiki/NAND_gate\#/media/File:CMOS_NAND.svg}}.
Refer to \cref{fig_nand} for the implementation.
\begin{figure}
  \centering
    \begin{tikzpicture}[x={(5mm,0mm)}, y={(0mm,-5mm)}]
  \draw[red] (0,-1) rectangle (28,14);

  \path (0,2) pic["$x_+$"] (xp) {I+};
  \path (0,4) pic["$x_-$"] (xm) {I-};
  \path (0,6) pic["$y_+$"] (yp) {I+};
  \path (0,13) pic["$y_-$"] (ym) {I-};

  \path (1,0) pic (S1p) {S+};
  \path (1,8) pic (S2m) {S-};
  \path (1,11) pic (S3p) {S+};

  \path (2.5,0) pic (W1p) {W+};
  \draw (S1p/S) -- (W1p/Ip/C);

  \path (2.5,8) pic (W2m) {W-};
  \draw (S2m/S) -- (W2m/Im/C);

  \path (2.5,11) pic (W3p) {W+};
  \draw (S3p/S) -- (W3p/Ip/C);





  \path (7.5,2) pic["$X_1$"] (X1p) {X+};
  \draw (xp/C) -- (X1p/Ip/C);
  \draw (xm/C) -- (X1p/Im/C);

  \path (7.5,6) pic["$S_2$"] (T2m) {T-};
  \draw (W2m/Om/C) -- (T2m/Im/C);
  \draw (yp/C) -- (T2m/Sp/C);

  \path (7.5,11) pic["$S_3$"] (T3p) {T+};
  \draw (W3p/Op/C) -- (T3p/Ip/C);
  \draw (ym/C) -- (T3p/Sm/C);

  \path (13,0) pic["$S_1$"] (T1p) {T+};
  \draw (W1p/Op/C) -- (T1p/Ip/C);
  \draw (X1p/Om/C) -- (T1p/Sm/C);

  \path (13,5) pic["$S_4$"] (T4m) {T-};
  \draw (T2m/Om/C) -- ++(0.5,0) -- ++(1,-1) -- (T4m/Im/C);
  \draw (X1p/Op/C) -- ++(0.5,0) -- ++(1,1) -- (T4m/Sp/C);

  \path (18,7) pic["$X_2$"] (X2m) {X-};
  \draw (T4m/Om/C) -- (X2m/Im/C);
  \draw (T3p/Op/C) -- ++(4.5,0) -- ++(2,-2) -- (X2m/Ip/C);

  \path (23,5) pic["$M$"] (Dp) {Dj+};
  \draw (X2m/Op/C) -- (Dp/I2p/C);
  \draw (T1p/Op/C) -- ++(2,0) -- ++(4,5) -- (Dp/I1p/C);

  \path (28,6) pic["$z_+$"] (zp) {O+};
  \draw (Dp/Op/C) -- (zp/C);

  \path (28,9) pic["$z_-$"] (zm) {O-};
  \draw (X2m/Om/C) -- (zm/C);
\end{tikzpicture}%
  \caption{\nand gate}
  \label{fig_nand}
\end{figure}
Notice the usage of switches means these gates are \emph{delay-sensitive}; that
is, the gate only operates correctly (i.e., computing the \nand function) if the
retarders have \emph{strictly greater delays} than the inputs $x$ and $y$.
In fact, for our construction we will need to instantiate this same
construction using \emph{varying} values for the retarders' delays
(but not their size as mentioned in \cref{sec_delays}).
This seems necessary in order to chain \nand gates in succession (since each
gate in a chain incurs a certain delay which must be compensated for in the next
gate down the chain).

In addition, notice that in principle \nand gates have \emph{variable} size as
their dimension depends on that of the three retarders,
As is the case for retarders, in the same embedding we insist on having all
\nand gates be of the same size.
We defer setting their dimensions to \cref{sec_setting_delays}; for now, it
suffices to keep in mind that \nand gates (and retarder elements) in the same
embedding only vary in their delay (and not their size).

\begin{claim}
  Assuming the retarders have larger delay than the input cables $x$ and $y$,
  the circuit on \cref{fig_nand} realizes a {\normalshape\nand} gate.
\end{claim}

\begin{proof}
  Consider first the case where both $x_+$ and $y_+$ are set.
  Since $x_-$ is not set, $X_1$ is consumed by $x_+$, turning $S_4$ on.
  In addition, since $y_+$ is set, $S_2$ is also turned on.
  Hence, using the assumption on the delay of the inputs, the negative source
  flows through $S_2$, $S_4$, and $X_2$ on to $z_-$.
  Since both the switches $S_1$ and $S_3$ remain open, the $z_+$ output is never
  set.
  Notice the crossings $X_1$ and $X_2$ are each used exactly once.

  Let now $x_-$ or $y_-$ (or both) be set.
  Then either $S_2$ or $S_4$ is open, which means $z_-$ is never set.
  As a result, $X_2$ is used at most once (namely in case $y_-$ is set).
  If $x_-$ is set, then $S_1$ is opened, thus allowing the positive source
  to flow on to $M$.
  The same holds if $y_-$ is set, in which case $M$ receives the positive
  source arriving from $S_3$.
  Hence, at least one positive signal will flow to the $M$ gate, causing $z_+$
  to be set eventually.
\end{proof}

\subsection{Cable Crossings}
\label{sec_cable_x}

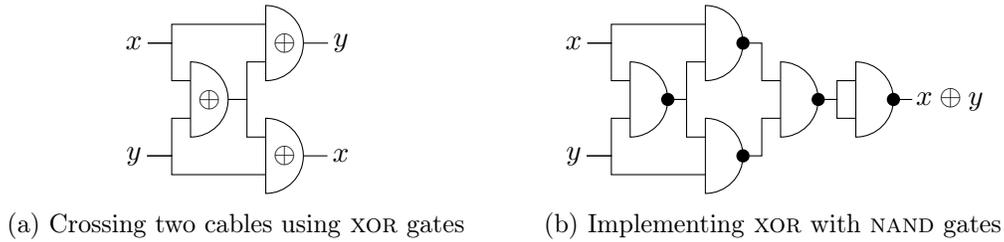
\begin{figure}
  \centering
  \hspace*{\fill}
  \begin{subfigure}[t]{0.4\textwidth}
    \centering
    \begin{tikzpicture}[x={(5mm,0mm)}, y={(0mm,-5mm)}]
  \path (-0.5,1) pic["x"] (I1) {I};
  \path (-0.5,4) pic["y"] (I2) {I};
  \path (1,2) pic (X1) {xor};
  \path (3,0.5) pic (X2) {xor};
  \path (3,3.5) pic (X3) {xor};
  \path (5,1) pic["y"] (O1) {I};
  \path (5,4) pic["x"] (O2) {I};

  \draw (I1/S) -- ++ (1,0) |- (X2/I1/C);
  \draw (I1/S) -- ++ (1,0) |- (X1/I1/C);
  \draw (I2/S) -- ++ (1,0) |-  (X3/I2/C);
  \draw (I2/S) -- ++ (1,0) |- (X1/I2/C);
  \draw (X1/O/C) -- ++ (0.5,0) |- (X2/I2/C);
  \draw (X1/O/C) -- ++ (0.5,0) |- (X3/I1/C);
  \draw (X2/O/C) -- (O1/S);
  \draw (X3/O/C) -- (O2/S);
\end{tikzpicture}
    \caption{Crossing two cables using \xor gates}
  \end{subfigure}
  \hspace*{\fill}
  \begin{subfigure}[t]{0.4\textwidth}
    \centering
    \begin{tikzpicture}[x={(5mm,0mm)}, y={(0mm,-5mm)}]
  \path (-0.5,1) pic["x"] (I1) {I};
  \path (-0.5,4) pic["y"] (I2) {I};
  \path (1,2) pic (N1) {nand};
  \path (3,0.5) pic (N2) {nand};
  \path (3,3.5) pic (N3) {nand};
  \path (5,2) pic (N4) {nand};
  \path (7,2) pic (N5) {nand};
  \path (9.5,2.5) pic["x\oplus y"] (O) {I};

  \draw (I1/S) -- ++ (1,0) |- (N2/I1/C);
  \draw (I1/S) -- ++ (1,0) |- (N1/I1/C);
  \draw (I2/S) -- ++ (1,0) |-  (N3/I2/C);
  \draw (I2/S) -- ++ (1,0) |- (N1/I2/C);
  \draw (N1/O/C) -- ++ (0.5,0) |- (N2/I2/C);
  \draw (N1/O/C) -- ++ (0.5,0) |- (N3/I1/C);
  \draw (N2/O/C) -- ++ (0.5,0) |- (N4/I1/C);
  \draw (N3/O/C) -- ++ (0.5,0) |- (N4/I2/C);
  \draw (N4/O/C) -- ++ (0.5,0) |- (N5/I1/C);
  \draw (N4/O/C) -- ++ (0.5,0) |- (N5/I2/C);
  \draw (N5/O/C) -- ++ (0.5,0) ;
\end{tikzpicture}
    \caption{Implementing \xor with \nand gates}
  \end{subfigure}
  \hspace*{\fill}
  \caption{Implementing cable crossings as in
  \cite{Goldschlager_1977_MPCVP_sigactn}}
  \label{fig:bitcrossing}
\end{figure}

There is a more or less well-known idea to cross to bits using three
\xor gates which can for example be found in the paper by
\textcite{Goldschlager_1977_MPCVP_sigactn}.
\cref{fig:bitcrossing} shows the idea.
This construction can be used in FA.
Because of the delays, there is not \emph{the} crossing gate, but a whole family
of them.
Depending on the position in the whole circuit layout, each crossing
needs \nand gates with specific builtin delays (which will be set in
\cref{sec_setting_delays}).

\section{Layer 4: Layout of a Whole Circuit}

Finally we describe one possibility to construct a finite rectangle of
cells $F$ of a FA containing the realization of a complete circuit, given
its description $B$.
The important point here is that, in order to produce $F$ from $B$, the
constructor only needs logarithmic space.
(Therefore the simplicity of the layout has precedence over any form of
\enquote{optimization}.)

\subsection{Arranging the Circuit in Tiles}

Let $C$ be the circuit that is to be embedded as an FA configuration
$F$. Letting $n$ be the length of inputs to $C$ and $m$ its number of
gates, notice we have an upper bound of $m$ on the circuit depth of
$C$. Without restriction, we may assume $m \ge n$, which also implies
an upper bound of $m+n = O(m)$ on the number of cables of $C$ (since
$C$ has bounded fan-in).
The logical gates of $C$ are denoted by $G_1,\dots,G_m$ and we assume that $G_i$
has number $n+i$ in description $B$ of $C$ (recall \cref{subsec:circuits}).

In the configuration $F$ we have cables $x_1,\dots,x_n$ originating from the
input gates as well as cables $g_1,\dots,g_m$ coming from (the embedding of) the
gates of $C$.
The $x_i$ and $g_i$ flow in and out of equal-sized \emph{tiles} $T_1,\dots,T_m$,
where in the $i$-th tile $T_i$ we implement the $i$-th gate $G_i$ of $C$.
The inputs to $T_i$ are $I_i = \{ x_1,\dots,x_n,g_1,\dots,g_{i-1} \}$ and its
outputs $O_i = I_i \cup \{ g_i \}$; hence $I_{i+1} = O_i$.

Recall that, unlike standard circuits, the behavior of our layer 3 circuits is
subject to spatial considerations, that is, to both gate placement and wire
length.
For the sake of simplicity, each tile is shaped as a square and all tiles are of
the same size.
In addition, the tiles are placed in ascending order from left to right and with
no space in-between.
The only objects in $F$ that lie outside the tiles are the inputs and output of
$C$ itself.
The inputs are placed immediately next to corresponding cables that go into
$T_1$ whereas the output is placed next to its corresponding wire $g_m$ at the
outgoing end of $T_m$.

\subsection{Layout for Tile \texorpdfstring{$i$}{i}}

\begin{figure}
  \centering
  \begin{tikzpicture}[x={(4mm,0mm)}, y={(0mm,-4mm)}]
  \draw[fill=facolorthree,draw=none] (0,0) rectangle (12.2,7.4);
  \draw[fill=facolorinput!50,draw=none] (0,7.6) rectangle (12.2,12);

  \draw (0,0) rectangle (12.2,12);
  \node at (6,1) {$T_i$};

  \foreach \y in {2,4,5,7} {
    \draw[-{Latex[]}] (-1,\y) -- ++(1,0);
    \draw[-{Latex[]}] (0,\y) -- ++(13,0);
  }

  \node[anchor=mid east] at (-1,2) {$x_1$};
  \node[anchor=mid west] at (13,2) {$x_1$};
  \node[anchor=mid east] at (-1,4) {$x_n$};
  \node[anchor=mid west] at (13,4) {$x_n$};
  \node[anchor=mid east] at (-1,5) {$g_1$};
  \node[anchor=mid west] at (13,5) {$g_1$};
  \node[anchor=mid east] at (-1,7) {$g_{i-1}$};
  \node[anchor=mid west] at (13,7) {$g_{i-1}$};

  \foreach \y in {2.7,3,3.3,5.7,6,6.3} {
    \draw[fill] (-0.3,\y) circle[radius=0.4pt];
    \draw[fill] (12.7,\y) circle[radius=0.4pt];
  }

  \draw (1.5,2) circle[radius=3pt];
  \draw[-{Latex[]}] (1.5,2) -- ++(5,7.5) -- ++(1,0);

  \draw (1.5,5) circle[radius=3pt];
  \draw[-{Latex[]}] (1.5,5) -- ++(3.5,5.5) -- ++(2.5,0);

  \draw (9.5,9) node[draw,anchor=north east,minimum size=8mm] (G) {$G_i$};
  \draw[-{Latex[]}] (G.east) -- ++(0.5,0) -- ++(2,-2) -- ++(1,0)
  node[anchor=mid west]  {$g_i$};

\end{tikzpicture}
  \caption{Overview of the tile $T_i$. The \emph{upper part} of the
    tile has green background, the \emph{lower part} has blue
    background.}
  \label{fig_tile}
\end{figure}
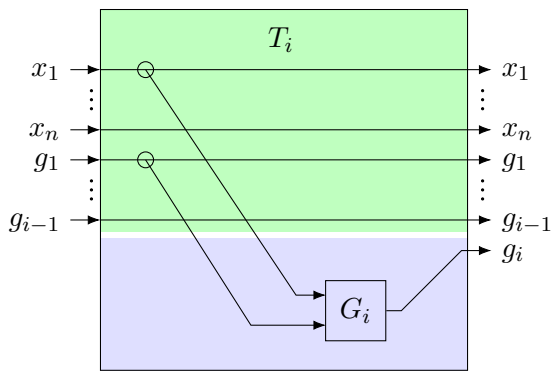

As depicted in \cref{fig_tile}, each tile is subdivided into two areas.
The \emph{upper part} contains the wires that pass through it, while the
\emph{lower part} implements the gate $G_i$ proper.

We give a broad overview of the process for constructing $T_i$.
First determine the numbers $y_1$ and $y_2$ of the inputs to $G_i$.
Then duplicate the bits on cables $y_1$ and $y_2$ (as in \cref{sec_bit_dupl})
and cross the copies over to the lower part of the tile.
These crossings require setting adequate delays, which will be adressed in the
next section.
(In case $y_1 = y_2$, duplicate the cable twice and proceed as otherwise
described.)
Next instantiate $G_i$ with a proper amount of delay (again, see the next
section) and plug in $y_1$ and $y_2$ as inputs into $G_i$.
Finally connect all inputs in $I_i$ as well as the output wire $g_i$ of $G_i$
to their respective outputs.
Notice the tile contains $O(m)$ crossings and thus also $O(m)$ \nand gates in
total.

\subsection{Choosing Suitable Delays for All Gates}
\label{sec_setting_delays}

The two details that remain are setting the dimensions and the delays for the
retarders in all \nand gates.
This requires certain care since we may otherwise end up running into a
chicken-and-egg problem:
The retarders' dimensions are determined by the required delays (in order to
have enough space to realize them); in turn, the delays depend on the
aforementioned dimensions (since the input wires in the \nand gates must be laid
so as to \enquote{go around} the retarders).

The solution is to assume we already have an upper bound $D$ on the maximum
delay in $F$.
This allows us to fix the size of the components as follows:
\begin{itemize}
  \item The retarders and \nand gates have side length $O(\sqrt{D})$.
  \item Each tile has side length $O(m \sqrt{D})$.
  \item The support of $F$ fits into a square with side length $O(m^2
  \sqrt{D})$.
\end{itemize}
With this in place, we determine upper bounds on the delays of the upper gates
in a tile (i.e., the gates in the upper part of the tile), then of the lower
gates $G_i$, then of the tiles themselves, and finally of the entire embedding
of $C$.
In the end we obtain an upper bound for the maximum possible delay in $F$.
Simply setting $D$ to be at least as large concludes the construction.

\paragraph{Upper gates.}
In order to set the delays of a \nand gate $G$ in a tile $T_i$, we first need an
upper bound $d_\mathrm{input}$ on the delays of the two inputs to $G$.
Suppose the origins $O_1$ and $O_2$ of these inputs (i.e., either a \nand gate
output or an input to $T_i$) have delay at most $d_\mathrm{origin}$.
Then certainly we have $d_\mathrm{input} \le d_\mathrm{origin} +
d_\mathrm{cable}$, where $d_\mathrm{cable}$ is the maximum of the cable
distances between either one of $O_1$ and $O_2$ and the switches they are
connected to inside $G$.
Due to the layout of a tile and since a \nand gate has $O(\sqrt{D})$ side
length, we know $d_\mathrm{cable}$ is at most $O(\sqrt{D})$.
Hence, if $G$ is in the $j$-th layer of $T_i$, then we may safely upper-bound
its delay by $d_i + (j+1) d_\mathrm{cable}$, where $d_i$ is the maximum over the
delays of the inputs to $T_i$.

\paragraph{Lower gates.}
Since there are $O(m)$ cables inside a tile, there are $O(m)$ cable crossings
and thus $O(m)$ \nand gates realizing these crossings.
Hence the inputs to the gate $G_i$ in the lower part of $T_i$ have delay at most
$d_i + O(m) \cdot d_\mathrm{cable} + O(m \sqrt{D})$, where the last factor is
due to the side length of $T$ (i.e., the maximum cable length needed to connect
the last of the upper gates with $G_i$).

\paragraph{Tiles.}
Clearly the greatest delay amongst the output cables of $T_i$ is that of $g_i$
(since every other cable originates from a straight path across $T_i$).
As we have determined in the last paragraph, at its output $g_i$ has delay
$d_{i+1} \le d_i + O(m \sqrt{D})$.
Since the side length of a tile is $O(m \sqrt{D})$, we may upper-bound the
delays of the inputs of $T_i$ by $i \cdot O(m \sqrt{D})$.

\paragraph{Support of $F$.}
Since there are $m$ tiles in total, it suffices to choose a maximum delay $D$ 
that satisfies $D \ge c m^2 \sqrt{D}$ for some adequate constant $c$ (that
results from the considerations above).
In particular, this means we may set $D = \Theta(m^4)$ independently of $C$.

\subsection{Constructor}
\label{subsec:constructor}

In this final section we describe how to realize a logspace constructor $R$
which, given a CVP instance consisting of the description of a circuit $C$ and
an input $x$ to it, reduces it to an instance as in \cref{main-thm}.
Due to the structure of $F$, this is relatively straightforward.

The constructor $R$ outputs the description of $F$ column for column.
(Computing the coordinates of an element or wire is clearly feasible in
logspace.)
In the first few columns $R$ sets the inputs to the embedded circuit according
to $x$.
Next $R$ constructs $F$ tile for tile.
To construct tile $T_i$, $R$ determines which cables are the inputs to $G_i$ and
constructs crossings accordingly.
To estimate the delays of each wire, $R$ uses the upper bounds we have
determined in \cref{sec_setting_delays}, which clearly are all computable in
logspace (since the maximum delay $D$ is polynomial in $m$).

Finally $R$ also needs to produce $y$ and $T$ as in the statement of
\cref{main-thm}.
Let $c_i$ be the cable of $T_m$ that corresponds to the output of the embedded
circuit $C$.
Then we let $y$ be the index of the cell next to the \pl wire of $c_i$ at the
output of $T_m$.
(Hence $y$ assumes a non-zero state if and only if $c_i$ contains a $1$, that
is, $C(x) = 1$.)
As for $T$, certainly setting it to the number of cells in $F$ suffices (since
a signal needs to visit every cell in $F$ at most once).

\section{Summary}
We have shown that, for fungal automata with update sequence $HV$, the
prediction problem is $\Ptime$-complete, solving an open problem of
\citeauthor{Goles_CUF_2021_pla} \cite{Goles_CUF_2021_pla}.

\printbibliography

\end{document}